\documentclass[twocolumn,aps,prl,showpacs,floatfix]{revtex4}
\usepackage{graphicx}
\usepackage{subfigure}
\usepackage{epsfig}
\usepackage{psfrag}

\makeatletter

\usepackage{verbatim}

\makeatletter

\input epsf

\def\be{\begin{equation}}
\def\ee{\end{equation}}
\def\ba{\begin{eqnarray}}
\def\ea{\end{eqnarray}}

\makeatother

\begin{document}

\title{Magnetic Excitations in the High $T_c$ Iron Pnictides}

\author{D.~X.~Yao and E.~W.~Carlson}

\affiliation{Department of Physics, Purdue University, West Lafayette, IN 47907}

\date{\today}

\begin{abstract}
  We calculate the expected finite frequency neutron scattering
    intensity based on the two-sublattice collinear antiferromagnet
    found by recent neutron scattering experiments as well as by
    theoretical analysis on the iron oxypnictide $LaOFeAs$.  We
    consider two types of superexchange couplings between Fe atoms:
    nearest-neighbor coupling $J_1$ and next-nearest-neighbor coupling
    $J_2$.  We show how to distinguish experimentally between
    ferromagnetic and antiferromagnetic $J_1$.  Whereas magnetic
    excitations in the cuprates display a so-called resonance peak at
    $(\pi,\pi)$ (corresponding to a saddlepoint in the magnetic
    spectrum) which is at a wavevector that is at least close to
    nesting Fermi-surface-like structures, no such corresponding
    excitations exist in the iron pnictides.  Rather, we find
    saddlepoints near $(\pi,\pi/2)$ and $(0,\pi/2)$ (and symmetry
    related points), which are not close to nesting the Fermi surfaces.

\end{abstract}
\maketitle

The recent discovery of superconductivity exceeding 50K in a new class
of materials holds tremendous potential for understanding the origin
of high temperature superconductivity. \cite{kamihara08,shan08,wen08,fang43k,zhao52k,zhao55k}
Similar to the cuprate superconductors, the iron pnictides also have a
layered structure, and display magnetism in the undoped parent
compound.  Both become superconducting upon doping.  And like the
cuprates, the transition metal layer is believed to play an important
role in the superconducting pairing.  On the other hand, the parent
compound of $LaOFeAs$ is a poor metal at room temperature, as opposed
to a correlated insulator as in the cuprates.

Initially band structure calculations suggested the materials are
nonmagnetic but close to a strong magnetic instability.\cite{singh08,kotliar08,xu08} However,
subsequent calculations have shown that the antiferromagnetic state
has lower energy than the nonmagnetic state because of Fermi surface
nesting.\cite{cao08,ma08,dongwang08}  In Ref.~\onlinecite{dongwang08}, a stripe-like
antiferromagnetic ground state was suggested based on strong nesting
effects. Recent neutron scattering experiments\cite{dai08} have shown
that the parent compound of $LaOFeAs$ is a long-range ordered
antiferromagnet with a type of spin stripe order ({\em i.e.}
unidirectional spin density wave).
However the magnetic moment was found to be $0.36(5) \mu_B$ per
iron, which is much smaller than the calculated value of $\sim2.3
\mu_B$ per iron. \cite{cao08,ma08,dongwang08}.  

\begin{figure}[t]
\begin{center}
\resizebox*{0.8\columnwidth}{!}{\includegraphics{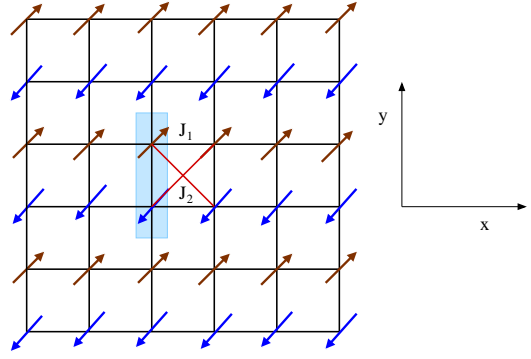}}
\end{center}
\caption{(Color online) Two-sublattice collinear antiferromagnet on the $Fe$-square lattice. Shaded region is the magnetic unit cell.}
\label{fig1}
\end{figure}

From an analysis of the superexchange interactions,
Ref.~\onlinecite{si08} suggested that the next-nearest-neighbor
interaction $J_2$ is antiferromagnetic (AFM), while the nearest-neighbor
interaction $J_1$ is ferromagnetic (FM). However a first-principles band
structure calculation predicts that the nearest-neighbor interaction
is also antiferromagnetic.\cite{tao08,ishibashi08} They predict that
$|J_2|$ is almost as twice large as $J_1$. In both cases, the
competition between $J_1$ and $J_2$ leads to a type of stripe-ordered
two-sublattice antiferromagnetic ground state (Fig~\ref{fig1}) when $|J_2/J_1|$ is
larger than the critical value.\cite{chandra90,shannon06} 
While the interactions $J_1$ and $J_2$ can compete, 
the uniaxial SDW considered in Fig.~\ref{fig1}
is a classical ground state
of the system, and it is thus not frustrated  in the sense of
having a macroscopic ground state degeneracy.

We use linearized spin wave theory to calculate the magnetic
excitations and sublattice magnetization for the two-sublattice
collinear antiferromagnet with nearest-neighbor superexchange coupling
$J_1$ and antiferromagnetic next-nearest-neighbor superexchange
coupling $J_2$.  We present results for ferromagnetic coupling $J_1$
as well as for antiferromagnetic coupling $J_1$.  (See
Fig.~\ref{fig1}.)  We find the results are quite different for the
two cases, so that comparing our calculations with future neutron
scattering results at finite frequency will be able to distinguish
these two cases.

The model Hamiltonian is described by the Heisenberg spin model on the square lattice
\begin{equation}
  H= J_1\sum_{{\langle ij \rangle}_{nn}}\mathbf{S}_i \cdot \mathbf{S}_j + J_2 \sum_{{\langle ij \rangle}_{nnn}}
\mathbf{S}_i \cdot \mathbf{S}_j
\end{equation} 
where $<ij>_{nn}$ and $<ij>_{nnn}$ mean the nearest-neighbor and next-nearest-neighbor spin pairs respectively. 
There are two spins in each unit cell, as shown in Fig.~\ref{fig1}. We
study the elementary excitations of the classical ground state of this model by using the
well-known Holstein- Primakoff boson method. The dispersion and intensities are calculated by
quantizing the classical spin waves.

We use Holstein-Primakoff bosons to quantize about the collinear antiferromagnetic
ground state found in recent neutron scattering.\cite{dai08}
\begin{equation}
  H=E_{Cl}+S\sum_{\mathbf{k}} [A_{\mathbf{k}}
  a_{\mathbf{k}}^+a_{\mathbf{k}}+\frac{1}{2}(B_{\mathbf{k}}
  a_{\mathbf{k}}^+a_{-\mathbf{k}}^++B_{\mathbf{-k}}^*
  a_{\mathbf{k}}a_{-\mathbf{k}}]
\end{equation}
where $E_{Cl}=-2J_2NS^2$ is the classical ground state energy and
\begin{eqnarray}
   A_{\mathbf{k}} &=& (4J_2+2J_1 \cos{k_x}),  \\
   B_{\mathbf{k}} &=& (2J_1 \cos{k_y}+4J_2 \cos{k_x}\cos{k_y}).
\end{eqnarray}
We can diagonalize the Hamiltonian using the Bogoliubov transformation
\begin{equation}
  b_{\mathbf{k}}=\cosh{\theta_{\mathbf{k}}a_{\mathbf{k}}}-\sinh{\theta_{\mathbf{k}}} a_{-\mathbf{k}}^+.
\end{equation}
The diagonalized Hamiltonian is
\begin{equation}
H=\sum_{\mathbf{k}}\omega(\mathbf{k})b_{\mathbf{k}}^+b_{\mathbf{k}}+E_{Cl}+E_{0}
\end{equation}
where $\omega(\mathbf{k})$ is the spin wave dispersion 
\begin{equation}
\omega(\mathbf{k})= S\sqrt{ A_{\mathbf{k}}^2- B_{\mathbf{k}}^2},
\end{equation}
and $E_0$ is the quantum zero-point energy correction
\begin{equation}
E_0=\frac{S}{2}\sum_{\mathbf{k}}(-A_{\mathbf{k}}+\omega(\mathbf{k})).
\end{equation} 
For $|J_1|=1$ and $J_2=2$, we get $E_0=-0.332NS$.

We find that there is only one spin wave band
\begin{eqnarray}
&&\omega(k_x,k_y)= \\
&&2S \sqrt{(2 J_2+ J_1 \cos{k_x})^2-(J_1 \cos{k_y}+2J_2 \cos{k_x} \cos{k_y})^2}. \nonumber
\end{eqnarray}
The associated spin wave velocities are 
\begin{eqnarray}
v_x &=& 2S\sqrt{-J_1^2+4J_2^2}, \\
v_y &=& 2S |J_1+2J_2|.
\end{eqnarray}
Notice that $v_x$ becomes imaginary for $|J_1| > 2|J_2|$,
indicating a change in the classical ground state configuration.

\begin{figure}[t]
{\centering
  \subfigure
  {\resizebox*{!}{0.7\columnwidth}{\LARGE{(a)}
  \includegraphics{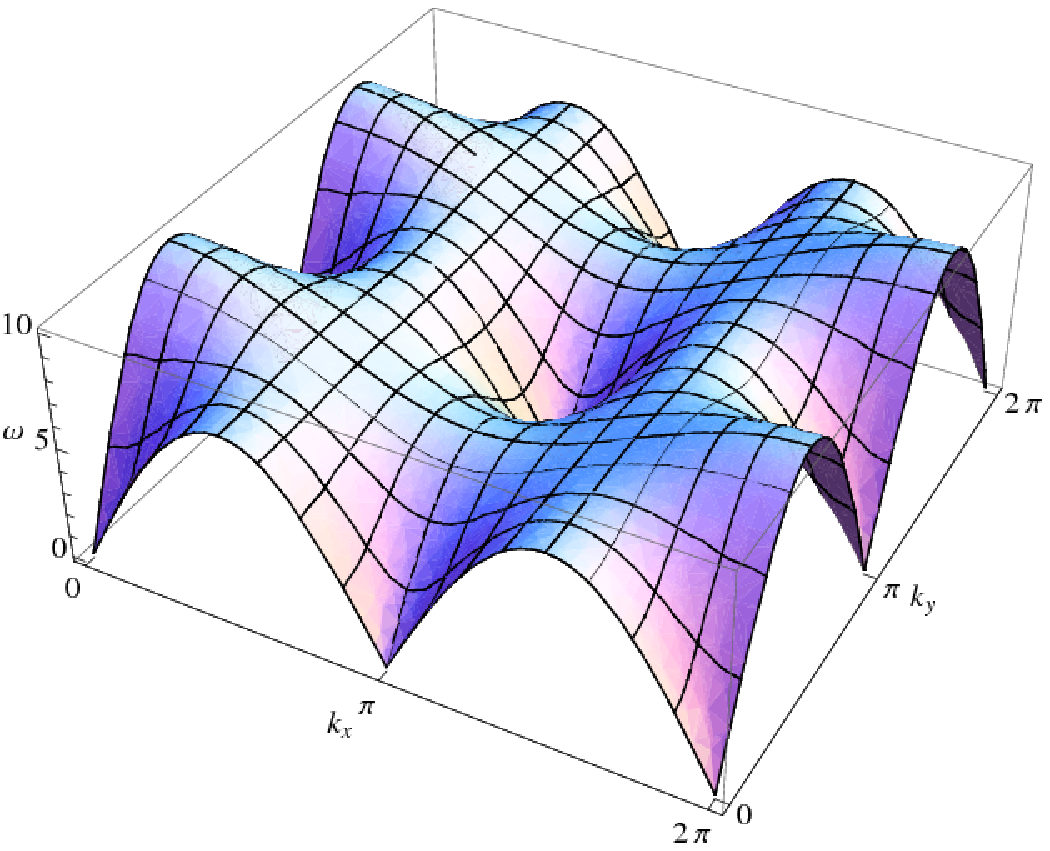}\label{3dafm2}}}
  \subfigure
  {\resizebox*{!}{0.7\columnwidth} {\LARGE{(b)}
\includegraphics{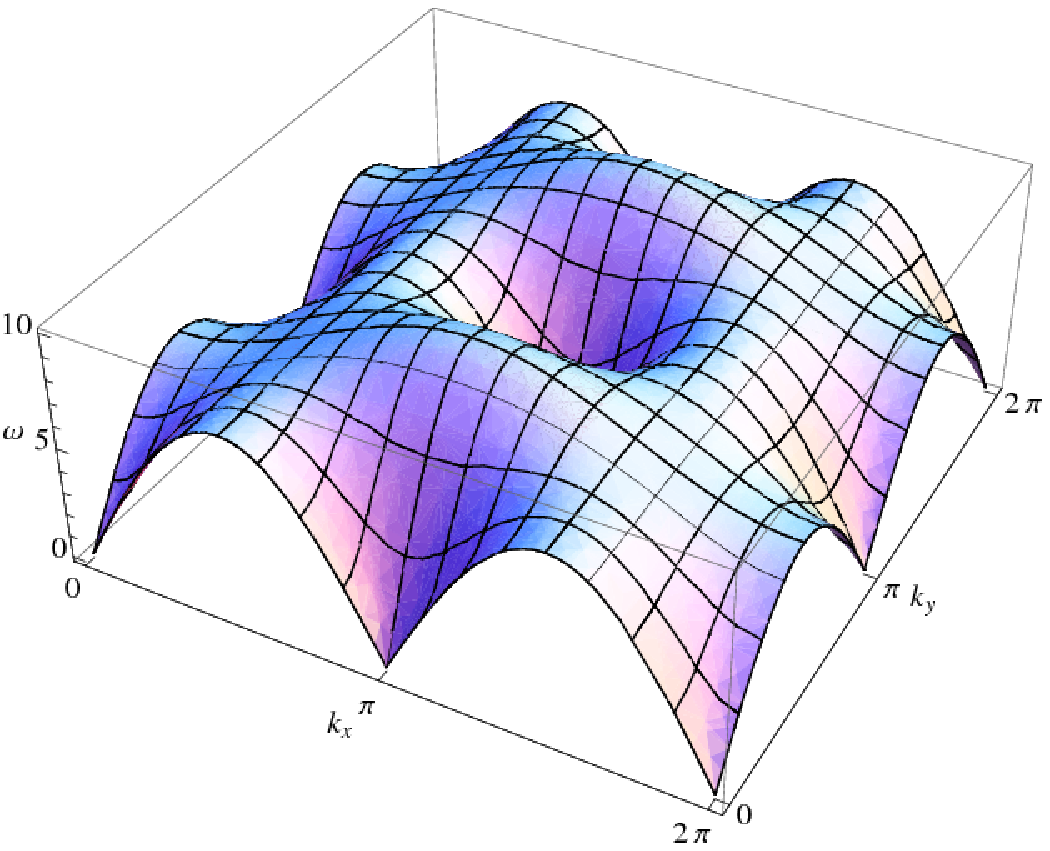}\label{3dfm2}}} 
  \par}
  \caption{(Color online) Spin-wave dispersion band for the two-sublattice collinear antiferromagnet
shown in Fig.~\ref{fig1}.  (a) Dispersion with both couplings antiferromagnetic.  
Here we have set $J_1 = 1$ (AFM) with $J_2 = 2$ (AFM). 
(b) Dispersion with ferromagnetic nearest neighbor coupling.
Here we have set $J_1 = -1$ (FM) with $J_2 = 2$ (AFM).}
\label{spinwave3d}
\end{figure}
Fig.~\ref{spinwave3d} shows the 
spin wave band with the nearest neighbor coupling both antiferromagnetic (Fig.~\ref{3dafm2})
and ferromagnetic (Fig.~\ref{3dfm2}).  
The presence of saddlepoints can be seen, and we will return to this point later.
In addition, because the $(\pi,\pi)$ point is a magnetic reciprocal lattice vector,
the dispersion must have $\omega \rightarrow 0$ at this point, although
as we will see there is no zero-frequency intensity associated with this part of the dispersion.
This precludes finite frequency weight at the $(\pi,\pi)$ point from this band.

We calculate the zero-temperature dynamic structure factor using
the same method, \cite{erica04,yao06a}
\begin{equation}
S(\mathbf{k}, \omega)=\sum_f \sum_{i=x,y,z} |<f|S^i (\mathbf{k})|0>|^2 \delta (\omega-\omega_f)~.
\end{equation}
Here $|0>$ is the magnon vacuum state and $|f>$ denotes the final
state of the spin system with excitation energy $\omega_f$.  $S^z$
does not change the number of magnons, contributing to the elastic
part of the structure factor. $\mathbf{S}^{x}(\mathbf{k})$ and
$\mathbf{S}^{y}(\mathbf{k})$ contribute to the inelastic dynamic
structure factor through single magnon excitations.

\begin{figure}[t]
{\centering
  \psfrag{w}{$\omega$}
  \psfrag{kx}{$k_x$}
  \psfrag{ky}{$k_y$}
  \subfigure
  {\resizebox*{!}{0.7\columnwidth} { \LARGE{(a)}
  \includegraphics{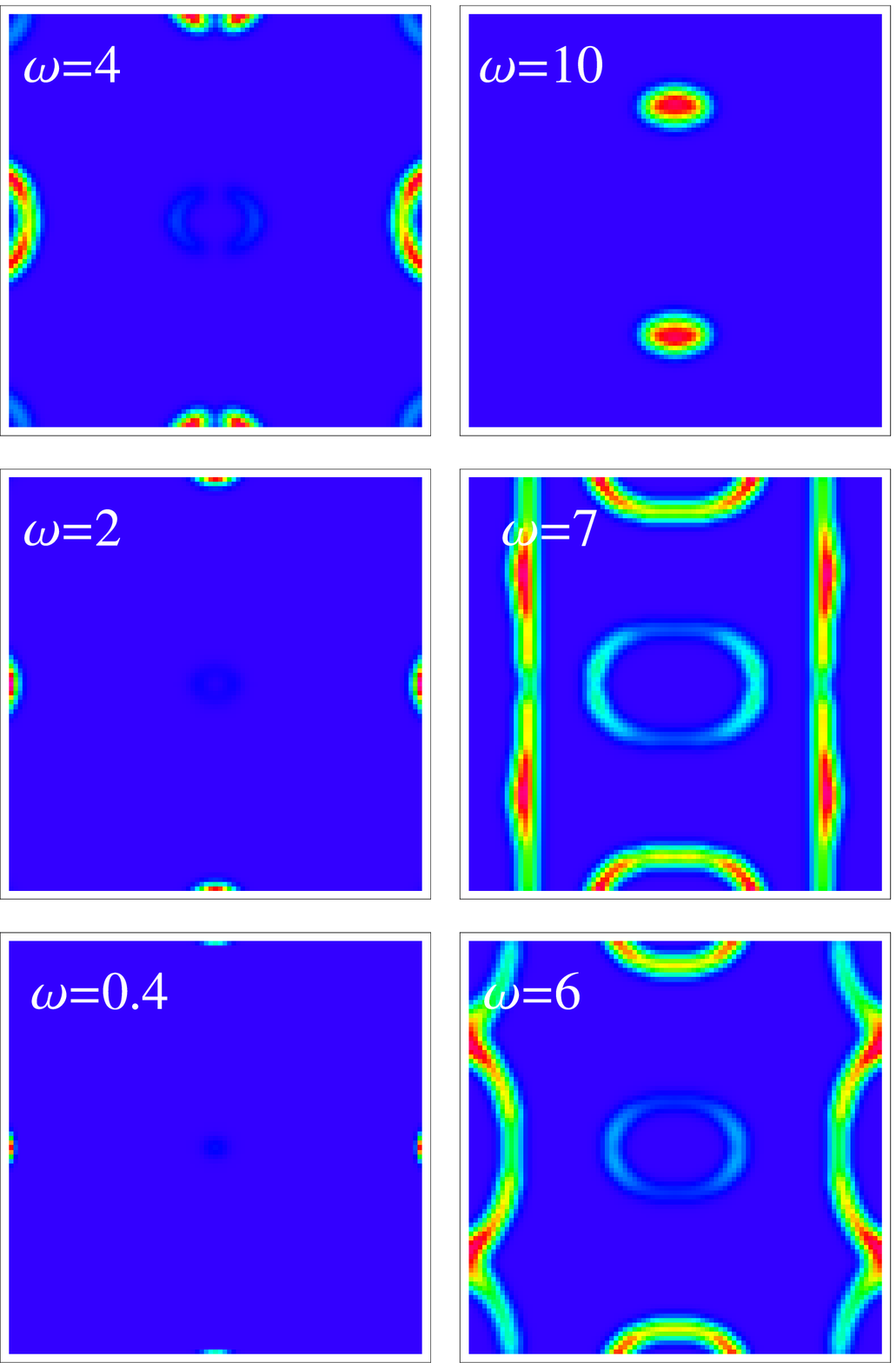}\label{ecuts1_fm}}}
  \subfigure
  {\resizebox*{!}{0.7\columnwidth} {\LARGE{(b)}
\includegraphics{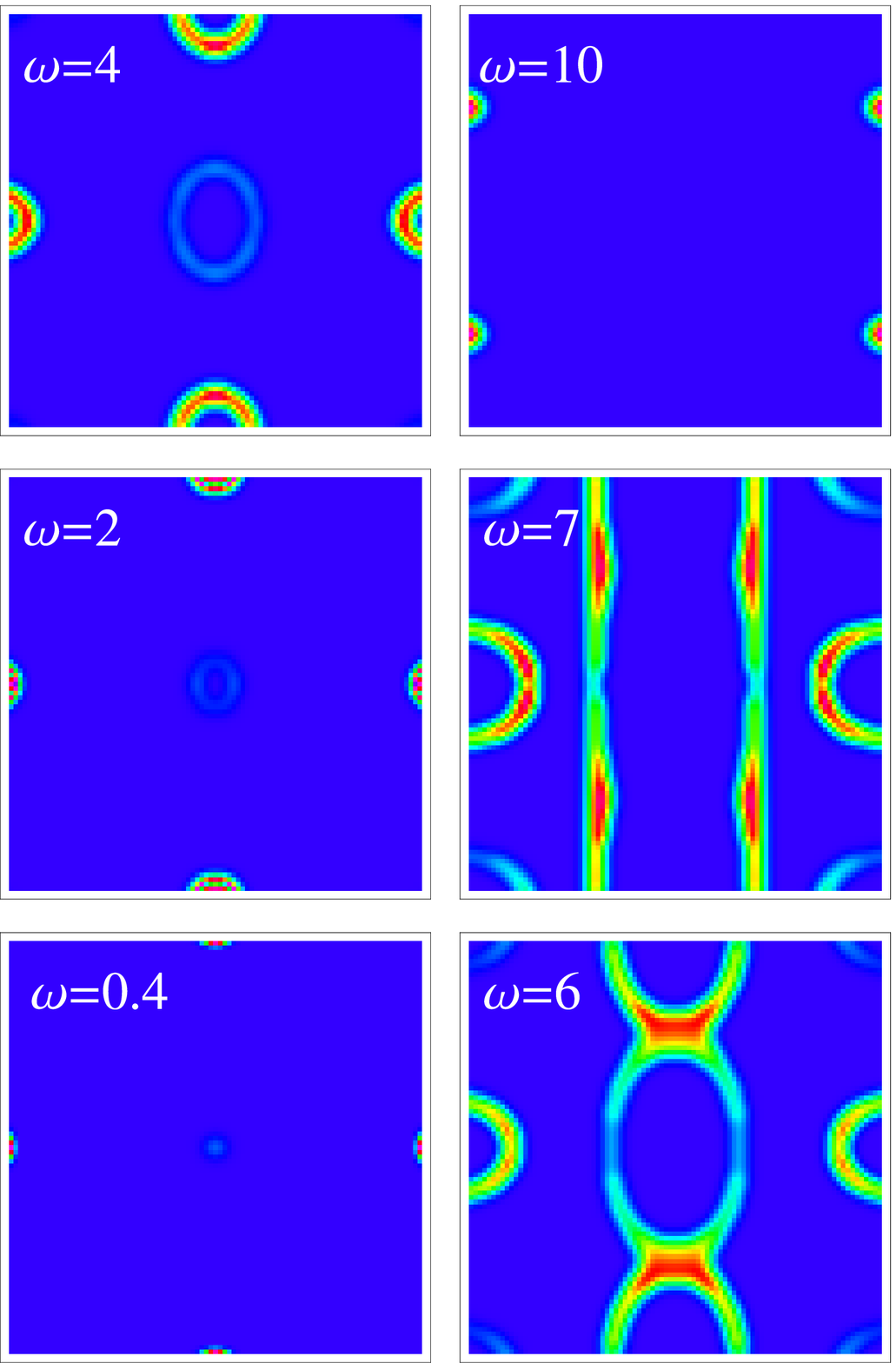}\label{ecuts1_af}}} 
  \par}
\caption{(Color online) Constant-energy cuts (untwinned) 
of the dynamic structure factor $S(\mathbf{k},\omega)$ 
for $J_2=2$ (AFM): (a) $J_1=-1$ (FM), (b) $J_1=1$ (AFM). The x-axis and y-axis correspond to $k_x$ and $k_y$
respectively
with the range $(0, 2\pi)$. 
We have integrated over an energy window of $\pm 0.2|J_1|S$.}
\label{untwinnedcuts}
\end{figure}

\begin{figure}[t]
{\centering
  \subfigure
  {\resizebox*{!}{0.7\columnwidth}{\LARGE{(a)}
  \includegraphics{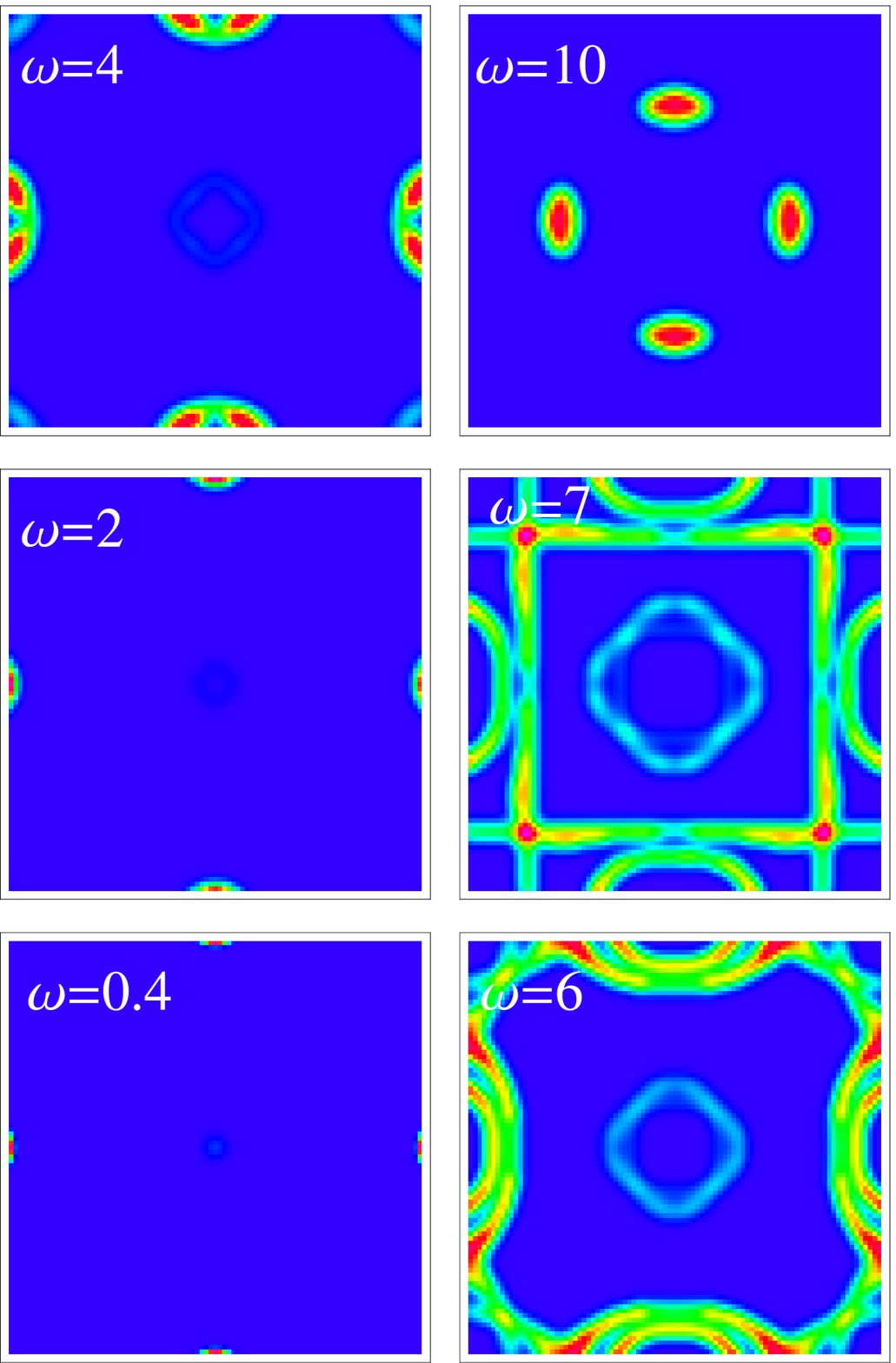}\label{ecuts2_fm}}}
  \subfigure
  {\resizebox*{!}{0.7\columnwidth}{\LARGE{(b)}\includegraphics{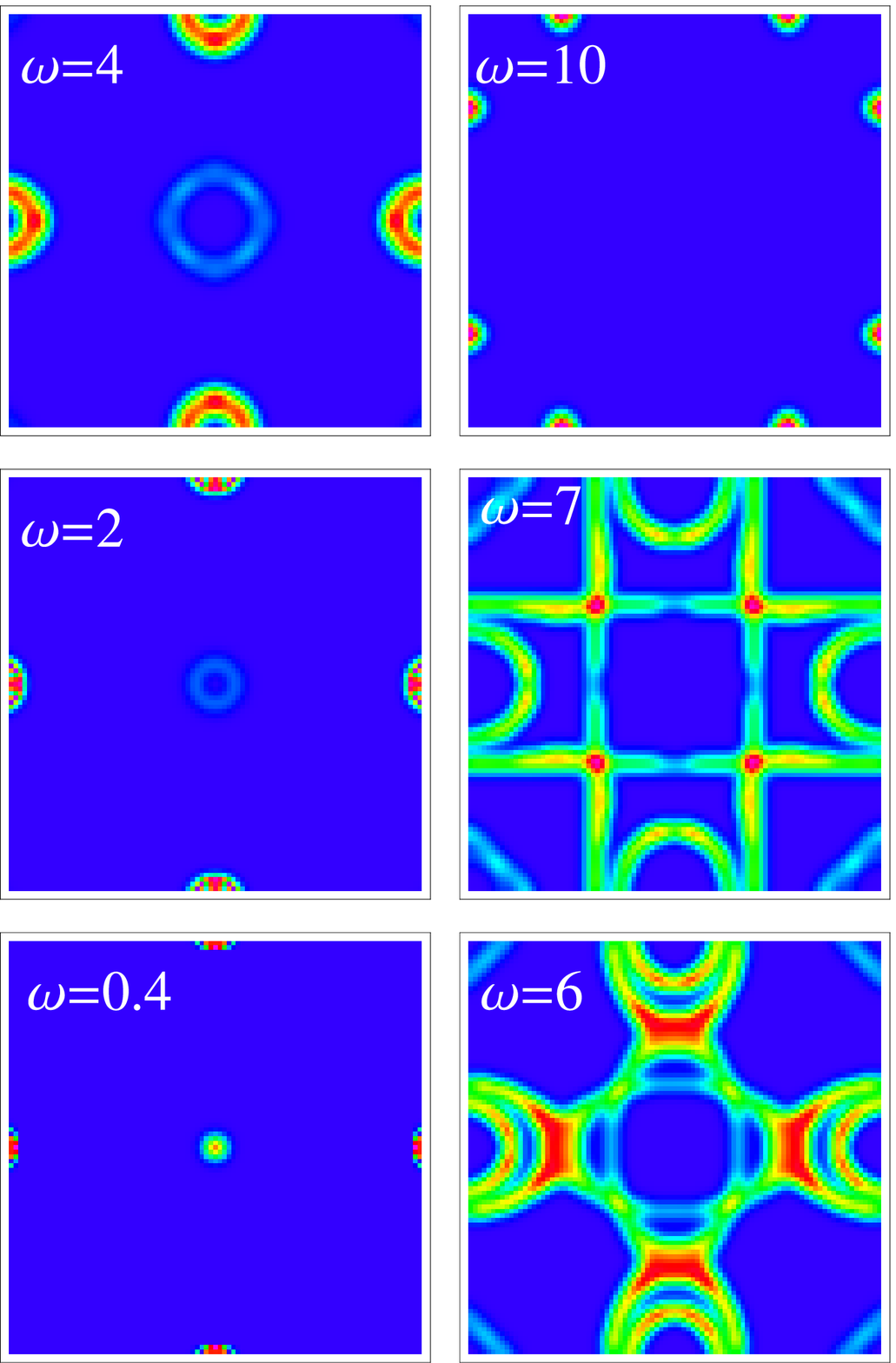}\label{ecuts2_af}}} 
  \par}
\caption{(Color online) Constant-energy cuts (twinned) 
of the dynamic structure factor $S(\mathbf{k},\omega)$ 
for $J_2=2$ (AFM): (a) $J_1=-1$ (FM), (b) $J_1=1$ (AFM). 
The x-axis and y-axis correspond to $k_x$ and $k_y$ respectively
with the range $(0, 2\pi)$. We have integrated over an energy window of $\pm 0.2|J_1|S$.}
\label{twinnedcuts}
\end{figure}

In Figs.~\ref{untwinnedcuts} and~\ref{twinnedcuts}, we show 
the expected neutron scattering intensity for 
constant energy cuts in $\mathbf{k}$-space.
We show our predictions from spin wave theory 
for both ferromagnetic and antiferromagnetic $J_1$.
Fig.~\ref{untwinnedcuts} shows the expected neutron scattering
intensity from a single domain of the magnetic order ({\em i.e.}
for an untwinned case), and Fig.~\ref{twinnedcuts} shows
the expected scattering intensity for the case where there is an
equal contribution from domains with both orientations of the
magnetic order ({\em i.e.} for a twinned case).

For ferromagnetic $J_1$, at low frequency, the 
strongest diffraction peaks are 
located at $(0, \pi)$.  (See Fig.~\ref{untwinnedcuts}.)
However more intensity weight shifts to $(\pi, 0)$ when $J_1$ is
antiferromagnetic. There is also a spin wave cone emerging from 
$(\pi,\pi)$, but the intensity is much weaker than 
the cones emanating from other magnetic reciprocal lattice vectors,
since zero frequency weight is forbidden at $(\pi,\pi)$
for the magnetic order we consider.
At high energy, the difference between 
ferromagnetic $J_1$ and antiferromagnetic $J_1$ becomes more apparent.
For example, for FM $J_1$, there are two strong spots along the $(\pi, k_y)$ direction, 
whereas for
AFM $J_1$, they are along the $(0, k_y)$ direction.  
In real materials, stripe order can be twinned due to, {\em e.g.}, a finite correlation length,
local disorder pinning, or crystal twinning.
Therefore we  show the twinned constant energy cut plots in
Fig.~\ref{twinnedcuts} for both FM and AFM nearest neighbor coupling $J_1$.

As can be seen from the dispersion in Fig.~\ref{spinwave3d},
there are saddlepoints in the spin wave excitation spectrum
at various points in $k$-space.  For the case of both couplings
antiferromagnetic, these occur at $(\pi/2,0)$ and $(\pi,\pi/2)$
and symmetry related points.  For ferromagnetic nearest neighbor coupling,
saddlepoints can be seen at $(0,\pi/2)$ along with weak saddlepoints
possible at $(\pi/2,0)$ and $(\pi/2,\pi)$ and symmetry related points.
The integrated intensity is generally large at such saddlepoints.
In the cuprates, there is a saddlepoint in the magnetic excitations at $(\pi,\pi)$ which has been
empirically connected to superconductivity, in that it increases in 
intensity at the onset of superconductivity, {\em i.e.} the ``resonance peak''.
There has been much discussion concerning  
this scattering phenomenon in the cuprates, particularly  because it is close to 
nesting vectors  for the corresponding Fermi surface.
However, in the case of the iron pnictides, the saddlepoints we
find here are quite far from any nesting vectors.

Experimentally, the magnetic moment per iron was found to be $0.36(5) \mu_B$,
 which is much
 smaller than the expected value of $\sim2.3 \mu_B$
per iron site.\cite{cao08,ma08,dongwang08}
The zero point energy of the spin waves reduces the sublattice
magnetization.  It was suggested in Ref.~\cite{si08}
that the competition between $J_1$ and $J_2$ may be 
responsible for the small moment observed in experiment.
The sublattice magnetization $m$ is defined as
\begin{equation}
m=<S_i^Z>=S-\Delta m,
\end{equation} 
 where $\Delta m$ is the deviation of sublattice magnetization from the saturation value,
\begin{eqnarray}
  \Delta m &=& <a_i^+a_i> \nonumber  \\
           &=& \sum_{\mathbf{k}}
           <a_{\mathbf{k}}^+a_{\mathbf{k}}>\nonumber \\
            &=& \frac{1}{2V_{\mathbf{k}}}\sum_{\mathbf{k}}
           [\frac{SA_{\mathbf{k}}}{\omega(\mathbf{k})}-1]+\frac{1}{V_{\mathbf{k}}}
           \sum_{\mathbf{k}} \frac{SA_{\mathbf{k}}}{\omega(\mathbf{k})}
           \frac{1}{e^{\beta \omega(\mathbf{k})}-1}\nonumber \\
           &=& \Delta m^{quantum} + \Delta m^{thermal}.
\end{eqnarray}
The first term $\Delta m^{quantum}$ comes from quantum zero point fluctuations.
The sceond term $\Delta m^{thermal}$ comes
from the classical thermal fluctuation, which is divergent at any
finite temperature in agreement with the Mermin-Wagner theorem.
(The very presence of the broken symmetry observed in experiment
implies that there is some finite coupling between planes, however weak.)

\begin{figure}[t]
\psfrag{J2}{$|J_2/J_1|$}
\psfrag{dm}{$\Delta m^{quantum}$}
\begin{center}
\resizebox*{0.8\columnwidth}{!}{\includegraphics{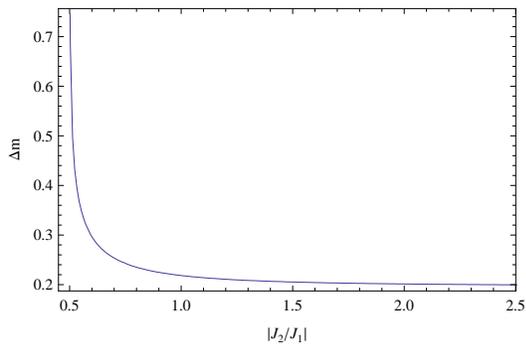}}
\end{center}
\caption{$|J_2/J_1|$ dependence of the reduction of the sublattice magnetization due to
 zero point energy of the spin waves. Here we have used $|J_1|=1$}
\label{dmag}
\end{figure}

Here we calculate $\Delta m^{quantum}$ by
\begin{equation}
  \Delta m^{quantum} =\frac{1}{2} \int_0^{2\pi}\int_0^{2\pi} \frac{dk_x}{2\pi} \frac{dk_y}{2\pi} \frac{SA_{\mathbf{k}}}{\omega(\mathbf{k})} -\frac{1}{2}.
\end{equation}
It is difficult to get the analytical form of the integral. Thus we
numerically calculate $\Delta m^{quantum}$. From the symmetry, the
above integral does not change when $J_1$ changes  sign. In
Fig.~\ref{dmag}, $\Delta m^{quantum}$ is plotted as a function of the
superexchange coupling ratio $|J_2/J_1|$. It is $S$ -independent. If
$S$ is in between $1$ and $\frac{3}{2}$, it will reduce the $m$ by
$13\%-20\%$. $\Delta m^{quantum}$ decreases with increasing $J_2/J_1$ because
stonger $J_2$ stablizes the two-sublattice collinear antiferromagnet
state.  This  deviation is not sufficient to explain the observed value of the
sublattice magnetization.

 In conclusion, we have used spin wave theory to calculate the
  magnetic excitations and sublattice magnetization for the
  two-sublattice collinear antiferromagnetic state of the new
  $La(O_{1-x}F_x)FeAs$ high-$T_c$ superconductors.  We have studied
  both ferromagnetic and antiferromagnetic nearest-neighbor coupling
  $J_1$ with antiferromagnetic next-nearest-neighbor coupling $J_2$.
  We calculate the predicted inelastic neutron scattering pattern
  based on spin wave theory. Comparison with future inelastic neutron
  scattering studies can be used to distinguish the sign of $J_1$.  We
  find that the sublattice magnetization can be reduced by the
  zero-point motion of spin waves, although not enough to account for
  the small moments observed in experiment.  In addition, we identify
  several saddlepoints in the magnetic excitation spectrum.  While
  magnetic excitations in these regions are expected to have extra
  intensity due to the saddlepoint structure, these corresponding
  wavevectors are not near nesting vectors of the Fermi surface.


\begin{acknowledgments}
We thank J. P. Hu, Y. L. Loh and A. Overhauser for helpful
discussions. D.~X.~Y. acknowledges support from Purdue
University. E. W. C is supported by Research Corporation. 

Note added: Some results from spin wave calculations have
also been reported by Ref.~\cite{hu08}. 

\end{acknowledgments}


\begin{thebibliography}{10}
\newcommand{\enquote}[1]{``#1''}

\bibitem{kamihara08}
Y.~Kamihara, T.~Watanabe, M.~Hirano, and H.~Hosono, J. Am. Chem. Soc. p. 3296
  (2008).

\bibitem{shan08}
L.~Shan, X.~Z. Y.~Wang, G.~Mu, L.~Fang, and H.-H. Wen, arXiv:0803.2405 .

\bibitem{wen08}
H.-H. Wen, G.~Mu, L.~Fang, H.~Yang, and X.~Zhu, Europhys. Lett. {\bf 82}, 17009
  (2008).

\bibitem{fang43k}
X.~H. Chen, T.~Wu, G.~Wu, R.~H. Liu, H.~Chen, and D.~F. Fang, arXiv:0803.3603 .

\bibitem{zhao52k}
Z.~A. Ren, J.~Yang, W.~Lu, W.~Yi, G.~C. Che, X.~L. Dong, L.~L. Sun, and Z.~X.
  Zhao, arXiv:0803.4283 .

\bibitem{zhao55k}
Z.~A. Ren, W.~Lu, J.~Yang, W.~Yi, Z.~C.~L. X.~L.~Shen, G.~C. Che, X.~L. Dong,
  L.~L. Sun, F.~Zhou, and Z.~X. Zhao, arXiv:0804.2053 .

\bibitem{singh08}
D.~Singh and M.~Du, arXiv:0803.0429 .

\bibitem{kotliar08}
K.~Haule, J.~H. Shim, and G.~Kotliar, arXiv:0803.1279 .

\bibitem{xu08}
G.~Xu, W.~Ming, Y.~Yao, X.~Dai, S.~Zhang, and Z.~Fang, arXiv:0803.1282 .

\bibitem{cao08}
C.~Cao, P.~J. Hirschfeld, and H.-P. Cheng, arXiv:0803.3236 .

\bibitem{ma08}
F.~Ma and Z.-Y. Lu, arXiv:0803.3286 .

\bibitem{dongwang08}
J.~Dong, H.~J. Zhang, G.~Xu, Z.~Li, G.~Li, W.~Z. Hu, D.~Wu, G.~F. Chen, X.~Dai,
  J.~L. Luo, Z.~Fang, and N.~L. Wang, arXiv:0803.3426 .

\bibitem{dai08}
C.~de~la Cruz, Q.~Huang, J.~W. Lynn, W.~R.~I. J.~Li, J.~L. Zarestky, H.~A.
  Mook, G.~F. Chen, J.~L. Luo, N.~L. Wang, and P.~Dai, Nature {\bf 453}, 899 (2008).

\bibitem{si08}
Q.~Si and E.~Abrahams, arXiv:0804.2480 .

\bibitem{tao08}
F.~Ma, Z.~Y. Lu, and T.~Xiang, arXiv:0804.3370 .

\bibitem{ishibashi08}
S.~Ishibashi, K.~Terakura, and H.~Hosono, arXiv:0804.2963 .

\bibitem{chandra90}
P.~Chandra, P.~Coleman, and A.~I. Larkin, Phys. Rev. Lett. {\bf 64}, 88 (1990).

\bibitem{shannon06}
N.~Shannon, T.~Momoi, and P.~Sindzingre, Phys. Rev. Lett. {\bf 96}, 027213
  (2006).

\bibitem{erica04}
E.~W. Carlson, D.~X. Yao, and D.~K. Campbell, Phys. Rev. B {\bf 70}, 064505
  (2004).

\bibitem{yao06a}
D.~X. Yao, E.~W. Carlson, and D.~K. Campbell, Phys. Rev. Lett. {\bf 97}, 017003
  (2006).

\bibitem{hu08}
C.~Fang, H.~Yao, W.-F. Tsai, J.~P. Hu, and S.~A. Kivelson, arXiv:0804.3843 .

\end{thebibliography}

\end{document}